\documentclass[%
nofootinbib,
nobibnotes,
 amsmath,amssymb,prl,twocolumn,
 aps,
]{revtex4-2}

 \usepackage{xcolor} 
 \usepackage{booktabs}
\usepackage{graphicx}
\usepackage{dcolumn}
\usepackage{bm}
\usepackage{slashed} 
\usepackage{tikz}
\usetikzlibrary{decorations.pathmorphing,decorations.markings}
\tikzset{
  photon/.style={decorate,decoration={snake,amplitude=0.6mm,segment length=2mm,pre length=0.3mm,post length=0.3mm}},
  fermion/.style={thick,postaction={decorate},decoration={markings,mark=at position 0.55 with {\arrow[scale=1.2]{stealth}}}},
  vtx/.style={circle,fill,inner sep=1pt},
  lab/.style={font=\small},
  num/.style={font=\scriptsize,text=gray},
}

\begin{document}


\title{\textbf{Rapid, High-Precision Numerical Evaluation of Multiloop Feynman Diagrams  } 
}%

\author{Richard Easther}
 \affiliation{%
 Department of Physics, Waipapa Taumata Rau $|$ University of Auckland, Private Bag 92019, Auckland, New Zealand.
}%

\date{\today}

\begin{abstract}
We numerically evaluate multiloop Feynman diagrams by using Sinc function approximations to transform the integrals into rapidly convergent, multidimensional sums. We compare the efficiency of this approach to Monte Carlo evaluations of electron $g-2$ terms. We show that the two strategies reach parity at precisions of  around 1\% for the full sets of three and four loop diagrams. Beyond this level Sinc function methods are dramatically faster -- computing 3-loop integrals to six digit precision takes minutes whereas Monte Carlo timings  map to years. We show that the method can evaluate individual 5-loop $g-2$ integrals and discuss extrapolation techniques that would further improve  performance. We note that the  method appears to be applicable to more general diagrams. 
\end{abstract}

\maketitle


The advent of quantum field theory \cite{Tomonaga:1946zz,Schwinger:1948yk,Feynman:1949zx} created the  need to evaluate the Feynman diagrams arising from its perturbative limit. This is a moving target:  rising experimental and theoretical sophistication requires tackling increasingly complex diagrams. The QED magnetic dipole moment, $g-2$, is emblematic of this evolution: it is now known analytically to four loops \cite{Schwinger:1948iu,Petermann:1957hs,Sommerfield:1957zz,Sommerfield:1958asm,Laporta:2017okg} but  the  experimental frontier is at five~\cite{Fan:2023edw,Aliberti:2025beg}. Diagrams that are unknown analytically are evaluated numerically    \cite{Aoyama:2014sxa,Aoyama:2012wj,
Aoyama:2012wk,Gabrielse:2006gg,Aoyama:2017uqe,
Aoyama:2020ynm,Aoyama:2019ryr,Volkov:2015pca,Volkov:2018jhy,Volkov:2019phy}. Monte Carlo methods are widely used for this \cite{Lepage:1977sw,Volkov:2018jhy} but their performance scales with the precision,~$\delta$, as $\delta^{-2}$, so that each  significant figure costs 100 times more  than the last.   This paper extends numerical approaches based on Sinc function expansions \cite{stenger1993numerical,Easther:1999qb,Easther:1999bs,Petrov:2000am}. (Note that the capitalised ``Sinc'' terminology follows Stenger \cite{stenger1993numerical}.) We evaluate all  $g-2$ diagrams up to 4-loops and a sample of 5-loop terms with a proof-of-concept implementation of the Sinc function method. We show  that it is competitive with Monte Carlo when $\delta \sim 10^{-2}$ for the overall loop correction terms and dramatically faster beyond it. 

The Sinc function method is applied by first moving the propagator denominators into exponentials with Schwinger parameters. Integrals over internal momenta become Gaussians and can be performed for any topology. The Sinc function expansion then converts the Schwinger integrals into rapidly convergent, multidimensional sums that  approximate the underlying diagrams. This  approximation is controlled by a parameter $h$, becoming exact as $h\rightarrow 0$.  The efficiency arises because the  sum approaches the exact result super-exponentially as $h$ decreases but the number of relevant terms only grows as a power. The sums converge  rapidly for larger $h$ and setting this parameter with a view to the desired precision greatly improves performance. 

This approach builds on earlier Sinc function evaluations of  simple scalar diagrams~\cite{Easther:1999qb}, 3-loop scalar diagrams \cite{Easther:1999bs} and 1-loop QED \cite{Petrov:2000am}. Related methods can evaluate the Fox functions arising from  Feynman diagram calculations~\cite{Passarino:2025cxi} and cosmological perturbation theory \cite{Slepian:2018vds}. Likewise, we draw heavily on previous analyses of the $g-2$ terms: we adopt Volkov's  ``forest'' implementation \cite{Volkov:2015pca,Volkov:2018jhy,Volkov:2019phy}  of the BPHZ scheme \cite{Bogoliubov:1957gp,Hepp:1966eg,Zimmermann1969}  and  benchmark against his results and those of  Kinoshita, Aoyama, Nio and  collaborators \cite{Aoyama:2014sxa,Aoyama:2012wj,Aoyama:2012wk,Gabrielse:2006gg,Aoyama:2017uqe,Aoyama:2020ynm,Aoyama:2019ryr}. Lastly, the heroic efforts of Laporta and others who developed analytic expressions for the 3- and 4-loop terms~\cite{Laporta:1996mq,Laporta:2017okg} allow us to check the validity of our results. 
Looking more widely, Sinc function methods  \cite{stenger1993numerical} appear in generic, rapidly convergent numerical integration schemes; the approach here is effectively a generalized trapezoidal integration \cite{trefethen2014exponentially}. Likewise, our handling of the propagator denominators  evokes Sinc function treatments of exponentially convergent approximations to $1/x$  \cite{Braess:2005ae}  and Poisson solvers \cite{Beylkin:2005oa,Exl:2016ae}.

Beyond the numerics, multiloop calculations in gauge theories with fermions have an irreducible level of algebraic complexity. However, agentic coding tools can handle this workload along with the implementation of the summation algorithm itself. The resulting code that produced our numerical results is available as {\sc sincFeyn}~\cite{sincFeyn}.

\section{The Method}

The electron and photon propagators are \cite{Peskin:1995ev}
\begin{equation}
   S_F(p) = \frac{i(\slashed{p}+m)}{p^2-m^2+i\epsilon}\, ,  \quad
D_F^{\mu\nu}(k) = \frac{-i\,g^{\mu\nu}}{k^2+i\epsilon}\, ,
 \end{equation}
as we work in the Feynman gauge \cite{Petrov:2000am}. Rewriting the denominators with Schwinger parameters,   
\begin{equation}
\frac{i}{p^2-m^2+i\epsilon} = \int_0^\infty ds\; e^{\,is(p^2-m^2+i\epsilon)} \, ,
\end{equation}
the integrals over internal momenta become Gaussians and are performed analytically.  The $g-2$ corrections are generated systematically along with their Dirac algebra and  regularized using Volkov's forest algorithm \cite{Volkov:2015pca,Volkov:2017xaq}.

\begin{figure}[tb]
    \centering
    \includegraphics[width=0.95\linewidth]{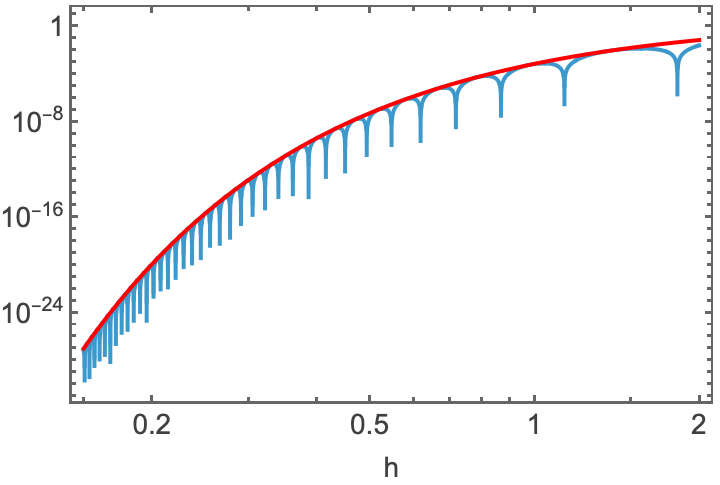}
        \caption{The exact fractional error in the Sinc expansion of the propagator as a function of $h$, with the error estimate from equation~\ref{eq:err} shown in red.   }
    \label{fig:properror}
\end{figure}

We then replace the Schwinger integrals with rapidly convergent sums using band-limited Sinc functions, also known as Shannon wavelets \cite{mallat2008wavelet}
\begin{equation}
S_k(h,x) = h \frac{\sin{\left[\frac{\pi (x-kh)}{h}\right]}}{{\pi (x-kh)}} = \frac{h}{2 \pi} \int_{-\frac{\pi}{h}}^{\frac{\pi}{h}}{e^{\pm i (x-kh)t}\,dt}.
\end{equation}
Orthogonality relations  \cite{Easther:1999qb,stenger1993numerical} allow definite integrals over the full real line to be approximated as 
\begin{equation}
\int_{-\infty}^{\infty} {f(z) dz } \approx h\sum_{k=-\infty}^{\infty} f(kh) \, ,
\label{sincintegral}
\end{equation}
and we use this with  the Schwinger integrals by setting $s=e^z$. After Wick-rotating, the  propagator denominator is  generically $1/X$ with $X>0$, and 
\begin{equation} \label{eq:propapprox}
    \frac{1}{X} \approx h \sum_{k=-\infty}^{\infty} 
    \exp{\left({k h} - e^{kh}X \right)} \, .
\end{equation}

There are generic error bounds for Sinc expansions \cite{stenger1993numerical} but for this case Poisson summation (or an appropriate contour integral) gives the right hand side of eq.~(\ref{eq:propapprox}) as  
\begin{equation}   
h\sum_{k=-\infty}^{\infty} e^{k h-X e^{kh}}
 =\sum_{n=-\infty}^{\infty}\Gamma \left(1+i \frac{2\pi  n}{h}\right)\frac{ X^{-i \frac{2\pi  n}{h}}}{X} \, .
\end{equation}
The $n=0$ term is precisely $1/X$ so the rest of the sum is an exact expression for the remainder. This series converges rapidly and applying Stirling's approximation (or identities for $\Gamma(1+i Y)$) to the $n\pm1$ terms gives an accurate estimate of the  fractional error when $h \lesssim 2$,  
\begin{equation} \label{eq:err}
    \mathrm{err} \approx \frac{4\pi}{\sqrt{h} } e^{-\pi^2 /h}\cdot \cos{(\cdots)} \, ,
\end{equation}
where $\cos{(\cdots)}$ denotes a cosine whose argument varies rapidly at small $h$.  Importantly, the  $\pi^2/h$ scaling means that $h$ is measured relative to a number that is roughly 10, so that $h\sim1$ is already ``small''.  The fractional error in the propagator is shown in Figure~\ref{fig:properror}.

\begin{figure}[tb]
    \centering
    \includegraphics[width=0.99\linewidth]{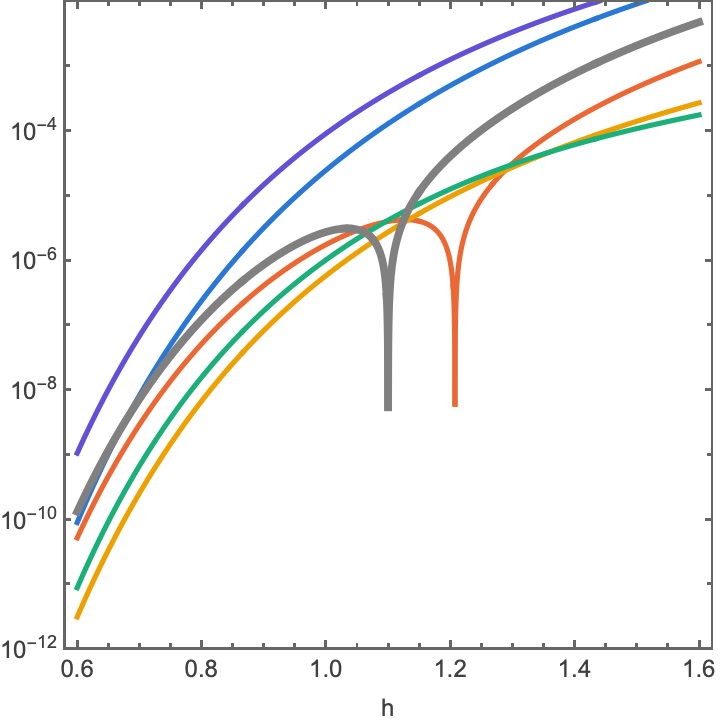}
    \caption{Fractional residuals for the 5 distinct 2-loop terms in Volkov's scheme as a function of $h$. Colors are given in Table~\ref{table:2loop}; the residual in the total $A_1^{(4)}$  is shown in gray. The self-energy  and $A_1^{(4)}$ residuals change sign in this $h$-range. }
    \label{fig:2loop}
\end{figure}

\definecolor{colcorner}{HTML}{2A78D6}
\definecolor{colselfenergy}{HTML}{EB6834}
\definecolor{colcrossed}{HTML}{1BAF7A}
\definecolor{colladder}{HTML}{EDA100}
\definecolor{colpitwo}{HTML}{6250D6}
\definecolor{colsum}{HTML}{808080} 
\begin{table}[tb]
\vspace{10pt}
\centering
\begin{tabular}{llcr}
\toprule
Graph & Exact value & Color & $10^{12} \Delta$ \\
\midrule
$\Pi_2$ insertion & $\frac{119}{36} - \frac{\pi^2}{3}$ & \textcolor{colpitwo}{\rule{1.2em}{0.6ex}} & $+71.27$ \\
Corner & $-\frac{55}{24} + \frac{\pi^2}{18} + \frac{\pi^2}{3}\log 2 - \frac{1}{2}\zeta(3)$ & \textcolor{colcorner}{\rule{1.2em}{0.6ex}} & $+4.92$ \\
Self-energy & $-\frac{1}{24} - \frac{\pi^2}{18}$ & \textcolor{colselfenergy}{\rule{1.2em}{0.6ex}} & $-3.69$ \\
Ladder & $\frac{11}{48} + \frac{\pi^2}{18}$ & \textcolor{colladder}{\rule{1.2em}{0.6ex}} & $-0.18$ \\
Crossed & $\frac{1}{6} + \frac{13\pi^2}{36} - \frac{5\pi^2}{6}\log 2 + \frac{5}{4}\zeta(3)$ & \textcolor{colcrossed}{\rule{1.2em}{0.6ex}} & $-0.55$ \\
\midrule
$A_1^{(4)}$ (sum) & $\frac{197}{144} + \frac{\pi^2}{12} - \frac{\pi^2}{2}\log 2 + \frac{3}{4}\zeta(3)$ & \textcolor{colsum}{\rule{1.2em}{0.6ex}} & $-9.43$ \\
\bottomrule
\end{tabular}
\caption{Two-loop terms, exact values, their color in Figure~\ref{fig:2loop} and their fractional residuals $\Delta$ ($\times 10^{12}$) at $h=0.55$; these were evaluated in seconds on a laptop. The weights for the mirrored diagrams are included. \label{table:2loop}}
\end{table}

\begin{figure}[tb]
    \centering
    \includegraphics[width=0.85\linewidth]{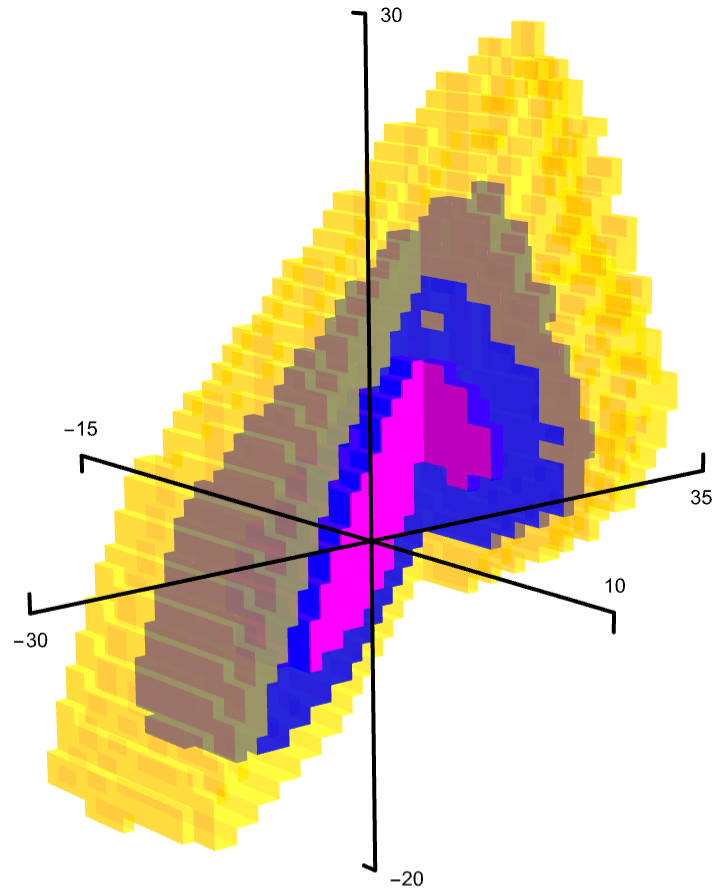}\\
\vspace{12pt}

        \caption{Terms in the sum with $h=1$ for the 2-loop self-energy diagram. Colors denote  points with absolute values $10^2$, $10^4$ and $10^6$ times smaller than the maximum; the sum is three dimensional after symmetry reductions.    }
    \label{fig:propshape}
\end{figure}

In earlier work~\cite{Easther:1999qb,Easther:1999bs,Petrov:2000am} the Sinc expansion was inserted first. By contrast, we begin with the Dirac algebra and generate the forest terms for each diagram -- all subsequent expressions are explicitly finite. We can always  set one Schwinger parameter to unity given that their overall volume can be rescaled  \cite{Cheng1987Expanding}, reducing the dimensionality by one. In addition, for  $g-2$ diagrams the momenta through the electron legs of the external photon vertex are necessarily identical because $q\rightarrow 0$, removing a second Schwinger parameter. Finally, many diagrams have symmetries (e.g. electron loops) where two Schwinger parameters can be rotated ($\lambda_{i,j}\rightarrow \lambda_i \pm \lambda_j)$ in a way that eliminates one of them, further reducing their order.  

The Sinc expansions are thus $D$-dimensional sums with $D\le 3L-2$, where $L$ is the loop order. As with the propagator, these sums approximate the underlying diagrams, becoming more accurate as $h$ decreases. Results for the five nontrivial 2-loop terms are shown in Figure~\ref{fig:2loop} and summarized in Table~\ref{table:2loop}.  The residuals between the converged sums and the exact results can cross zero but do so  less frequently than with a single propagator.  

The efficient evaluation of the sums is critical to the practical value of the algorithm. We auto-generate kernels for each diagram that are summed by a common driver algorithm. The kernels can be large (megabytes at four or five loops) and are processed with FORM \cite{Davies:2026cci} to maximize their computational efficiency. As well as the Sinc parameter {\tt{h}} the code takes inputs {\tt{eps}} and {\tt{N}} which set the minimum size of the terms to be kept (this must be substantially smaller than the expected accuracy of the sum) and an upper bound on the lattice parameters, which  must  be larger when {\tt{eps}} is smaller. 

Truncation error is monitored by keeping totals for $10\times {\tt{eps}}$ and $100\times {\tt{eps}}$; an estimate of the contribution from  ``pruned'' terms can be added to the final result. The sum is structured   to minimize round-off error. The impact of  the choice of $h$ is assessed empirically by evaluating sums at several values of this parameter, and via generic scaling relations.  The summand for the 2-loop ``self-energy'' term is shown in Figure~\ref{fig:propshape}; it is three dimensional after symmetries are imposed and its ``tail'' is longest in the negative $k$ directions.

The full, scheme-independent 2-loop result is  well-known \cite{Petermann:1957hs,Sommerfield:1957zz} but  individual terms depend on the  subtraction method and we are unaware of analytic forms for the corner and self-energy diagrams in the forest scheme. With $h=0.55$  2-loop diagrams take a few seconds on a laptop and deliver roughly ten digits of precision. These diagrams can be computed to  higher accuracy  with lower values of $h$ with quad precision arithmetic. With these results the PSLQ algorithm \cite{Ferguson:1999pslq,mpmath} identifies the missing analytic expressions which are included in  Table~\ref{table:2loop}.

\section{Numerical Performance}

The implementation of {\sc sincFeyn} \cite{sincFeyn} is heuristic, rather than built on a first-principles approach. The summands are approximately unimodal in magnitude but with oscillating signs, so there can be significant cancellations within an evaluation. In general, diagrams with smaller $\langle f\rangle /\langle|f| \rangle$  (where f is the summand and $\langle \cdots \rangle$ is an average) are more costly at fixed relative accuracy, as is the case for the Monte Carlo integrals \cite{Volkov:2019phy}. In addition, while  the regularized summands are finite term-by term they are often expressed as the differences of large numbers. These can require quad precision, slowing their evaluation -- we tune the ``fallback'' threshold to minimize these occurrences while preserving accuracy. 

For $D\le 10$ we start with an initial survey of the summand at steps of $2h$. This reduces the number of terms by $\sim 2^D$, making it quick to explore.   We store the points above the {\tt{eps}} threshold (checking for terms that are ``accidentally'' close to zero, given the oscillating sign) and then perform the full sum  over the $2^D$ points in each of the boxes. This avoids conditionals, aiding parallelization but leading to the evaluation of many terms exponentially smaller than the final result. However, the rising complexity of the summands with $D$ means that a ``top down'' summation becomes faster when $D\ge 11$; these diagrams first arise at five loops. 

There are 72, 891 and 12,672 3-, 4- and 5- loop diagrams (respectively)  that contribute to $g-2$. In particular 50, 518 and 6,354 diagrams without lepton loops yield 28, 269 and 3,213 unique integrals. We have evaluated all the $g-2$ diagrams for $L\le4$ over a range of $h$ values and the no-lepton-loop integrals account for most of the cost.  We map speeds using the nominal peak performance of the hardware platforms.  Monte Carlo timings were performed on 1.4 TFLOP/s  \cite{Volkov:2018jhy,nvidia2014teslak80} and 7.8 TFLOP/s \cite{Volkov:2019phy,nvidia2017teslav100} GPUs and are mapped via $T\sim \delta^{-2}$, ignoring possible non-Gaussian tails. Sinc evaluations used a 10 TFLOP/s GPU \cite{nvidia2020a100} and a desktop with approximately 10\% of its performance. None of our conclusions depend on the detailed timings.

\begin{figure}[tb]
    \centering
    \includegraphics[width=0.99\linewidth]{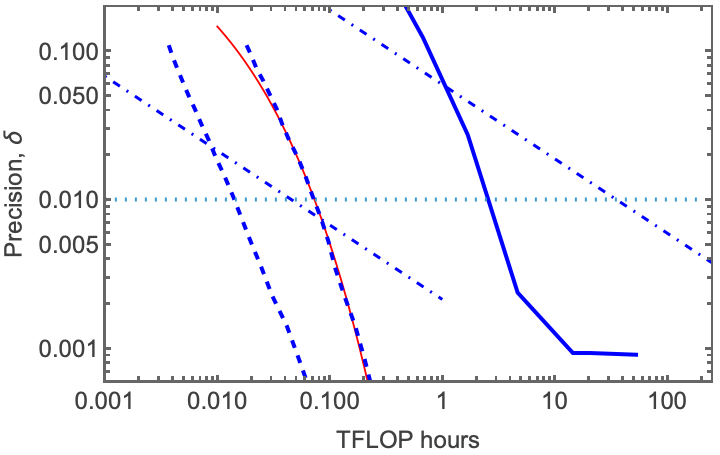} 
    \caption{ Timing for 3- (dashed) and 4-loop (solid) no-lepton-loop diagrams. The dotted line denotes 1\% precision, the dot-dashed lines are extrapolated Monte Carlo timings \cite{Volkov:2018jhy}. 3-loop results are shown with fixed (slower) and varying  {\tt eps} and a fit to equation~\ref{eq:timeaccuracy} (thin line) for the fixed {\tt eps} results. }
    \label{fig:times}
\end{figure}

Figure~\ref{fig:times} shows that the Sinc function method  matches  extrapolated Monte Carlo timings when $\delta\sim10^{-2}$ for 3- and 4-loop corrections and is dramatically faster at higher precision.  Figure~\ref{fig:times} gives 3-loop results for both fixed {\tt eps} (chosen to match the smallest value of $h$) and a varying {\tt eps} which is matched to the $h$-dependent truncation error. At 4-loops we only show varying {\tt eps} timings. Residuals can cross zero or have near-cancellations, giving ``accidental'' accuracy boosts. This leads to the knee in the 4-loop times; sums over gauge-invariant subgroups \cite{Laporta:2017okg,Volkov:2017xaq} of no-lepton loop diagrams are better behaved.

At fixed {\tt eps} the number of relevant points and thus the computational cost $T$ can be expected to scale as $(1/h)^{D_{\mathrm{eff}}}$. Equation~\ref{eq:err} suggests  $\log{(1/\delta)} \sim c/h$, giving
\begin{equation} \label{eq:timeaccuracy}
    T \sim a \log{\left(\frac{1}{\delta}\right)}^{D_{\mathrm{eff}}} \, .
\end{equation}
This fits the fixed {\tt eps} results with $D_{\mathrm{eff}} \approx 2.24$. This number depends on the implementation of the summations as well as the overall strategy, since {\tt eps} may not be a sharp cut, but it is clearly vastly different from the Monte Carlo scaling. 

The ``break-even point'' between Monte Carlo and the Sinc function method for the 3- and 4-loop no-lepton-loop terms is  1\% in the total, not 1\% in the individual diagrams. These must be evaluated more accurately; the error in the sum of $N$  terms is  roughly $\sqrt{N}$ larger than the per-term error and $\sqrt{N}\sim 60$ at 5 loops. Estimates from single diagram evaluations suggest that the 1\% threshold also holds at five loops. The current best value at this level is the  product of major  campaigns \cite{Volkov:2024yzc,Aoyama:2024aly}. Since their uncertainty is on the order of 1\%, improving it with Sinc methods would still be costly. 

\begin{figure}[tb]
    \centering
    \includegraphics[width=0.99\linewidth]{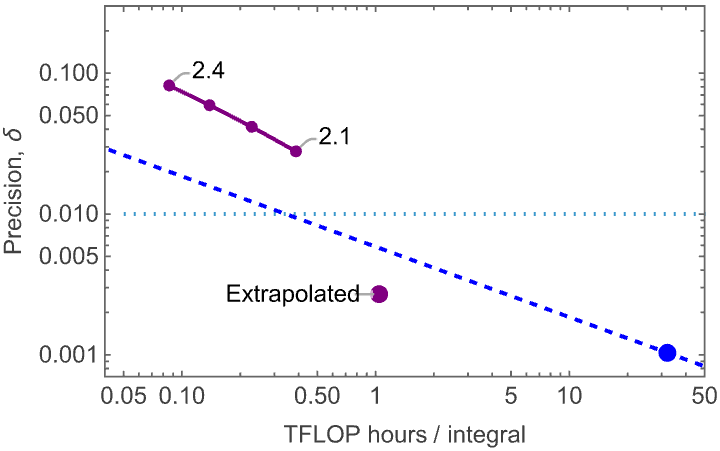} 
    \caption{ Average times and accuracies (relative to Monte Carlo evaluations) for a sample of 5 loop diagrams. Dashed: scaled Monte Carlo runtimes \cite{Volkov:2019phy}; Dot: actual runtime; Solid: Sinc function results (with $h$), and extrapolated value.  }
    \label{fig:times5}
\end{figure}

That said, the smooth relationship between the accuracy of Sinc representations and $h$ may make it possible to fit an empirical error function (again motivated by equation~\ref{eq:err}) to cheaper evaluations at $h\sim 2$ and then accurately infer the value at lower $h$.  Figure~\ref{fig:times5} illustrates  a simple implementation of this strategy which substantially improves the cost/performance relationship for a sample of  5-loop diagrams.

\section{Discussion and Prospects}
We have demonstrated that the Sinc function methods can provide a viable alternative to Monte Carlo evaluations of Feynman diagrams. In particular, testing against  known analytic values for  3 and 4-loop contributions to $g-2$  shows vast performance gains at higher accuracies. 

Sinc function methods can evaluate 5-loop $g-2$ diagrams but a na{\"\i}ve calculation of the full set of integrals  would still be very expensive. However, the intrinsic accuracy of Sinc function expressions depends smoothly on the parameter $h$ and we can estimate results at smaller $h$ by fitting to much cheaper evaluations at larger $h$. 
This would need a careful analysis of the $h$-dependence of the Sinc function expressions and should account for potential zero-crossings by the residuals, as seen in Figure~\ref{fig:times}. We could also directly evaluate the derivatives of the sums with respect to $h$ to form a Taylor series. This is  a clear avenue for future work.

Separately, this method appears to be compatible with dimensional regularization and diagrams with non-zero external momenta.  Moreover, early investigations suggest that it can accelerate some quasi-Monte Carlo evaluations in tools like {\sc pySecDec}~\cite{Borowka:2017smv}. These topics likewise represent interesting opportunities.

\vfill

\begin{acknowledgments}
I thank Mark Mueller, Washington Taylor and Yourong Wang for useful discussions. Sergey Volkov and Stefano Laporta provided insightful commentary on earlier drafts. I acknowledge the Waipapa Taumata Rau $|$ University of Auckland Centre for eResearch for  computational resources and I am  very grateful to Nidhi Gowdra for supporting the GPU runs. Finally, despite the passage of time, this work owes much to conversations with the late Gerald Guralnik. 
\end{acknowledgments}
\newpage

\section{AI Statement}
The manuscript has no AI-generated content; however {\sc sincFeyn} was entirely constructed via high-level prompts to Anthropic's Claude Opus 5.5. AI input included  design and implementation suggestions stemming from an open-ended request to replicate the calculations in Refs~\cite{Easther:1999bs,Easther:1999qb,Petrov:2000am} and to consider their multiloop extension in QED.  Claude was also used to manage the analysis of the results. However, the overall strategy and detailed summation approach was controlled by user input. The match between the output of  {\sc sincFeyn} and known results is a key verification step and parts of the code have been replicated externally.

 
%


\begin{thebibliography}{49}%
\makeatletter
\providecommand \@ifxundefined [1]{%
 \@ifx{#1\undefined}
}%
\providecommand \@ifnum [1]{%
 \ifnum #1\expandafter \@firstoftwo
 \else \expandafter \@secondoftwo
 \fi
}%
\providecommand \@ifx [1]{%
 \ifx #1\expandafter \@firstoftwo
 \else \expandafter \@secondoftwo
 \fi
}%
\providecommand \natexlab [1]{#1}%
\providecommand \enquote  [1]{``#1''}%
\providecommand \bibnamefont  [1]{#1}%
\providecommand \bibfnamefont [1]{#1}%
\providecommand \citenamefont [1]{#1}%
\providecommand \href@noop [0]{\@secondoftwo}%
\providecommand \href [0]{\begingroup \@sanitize@url \@href}%
\providecommand \@href[1]{\@@startlink{#1}\@@href}%
\providecommand \@@href[1]{\endgroup#1\@@endlink}%
\providecommand \@sanitize@url [0]{\catcode `\\12\catcode `\$12\catcode
  `\&12\catcode `\#12\catcode `\^12\catcode `\_12\catcode `\%12\relax}%
\providecommand \@@startlink[1]{}%
\providecommand \@@endlink[0]{}%
\providecommand \url  [0]{\begingroup\@sanitize@url \@url }%
\providecommand \@url [1]{\endgroup\@href {#1}{\urlprefix }}%
\providecommand \urlprefix  [0]{URL }%
\providecommand \Eprint [0]{\href }%
\providecommand \doibase [0]{https://doi.org/}%
\providecommand \selectlanguage [0]{\@gobble}%
\providecommand \bibinfo  [0]{\@secondoftwo}%
\providecommand \bibfield  [0]{\@secondoftwo}%
\providecommand \translation [1]{[#1]}%
\providecommand \BibitemOpen [0]{}%
\providecommand \bibitemStop [0]{}%
\providecommand \bibitemNoStop [0]{.\EOS\space}%
\providecommand \EOS [0]{\spacefactor3000\relax}%
\providecommand \BibitemShut  [1]{\csname bibitem#1\endcsname}%
\let\auto@bib@innerbib\@empty
\bibitem [{\citenamefont {Tomonaga}(1946)}]{Tomonaga:1946zz}%
  \BibitemOpen
  \bibfield  {author} {\bibinfo {author} {\bibfnamefont {S.}~\bibnamefont
  {Tomonaga}},\ }\bibfield  {title} {\bibinfo {title} {{On a relativistically
  invariant formulation of the quantum theory of wave fields}},\ }\href
  {https://doi.org/10.1143/PTP.1.27} {\bibfield  {journal} {\bibinfo  {journal}
  {Prog. Theor. Phys.}\ }\textbf {\bibinfo {volume} {1}},\ \bibinfo {pages}
  {27} (\bibinfo {year} {1946})}\BibitemShut {NoStop}%
\bibitem [{\citenamefont {Schwinger}(1948{\natexlab{a}})}]{Schwinger:1948yk}%
  \BibitemOpen
  \bibfield  {author} {\bibinfo {author} {\bibfnamefont {J.~S.}\ \bibnamefont
  {Schwinger}},\ }\bibfield  {title} {\bibinfo {title} {{Quantum
  electrodynamics. I A covariant formulation}},\ }\href
  {https://doi.org/10.1103/PhysRev.74.1439} {\bibfield  {journal} {\bibinfo
  {journal} {Phys. Rev.}\ }\textbf {\bibinfo {volume} {74}},\ \bibinfo {pages}
  {1439} (\bibinfo {year} {1948}{\natexlab{a}})}\BibitemShut {NoStop}%
\bibitem [{\citenamefont {Feynman}(1949)}]{Feynman:1949zx}%
  \BibitemOpen
  \bibfield  {author} {\bibinfo {author} {\bibfnamefont {R.~P.}\ \bibnamefont
  {Feynman}},\ }\bibfield  {title} {\bibinfo {title} {{Space - time approach to
  quantum electrodynamics}},\ }\href {https://doi.org/10.1103/PhysRev.76.769}
  {\bibfield  {journal} {\bibinfo  {journal} {Phys. Rev.}\ }\textbf {\bibinfo
  {volume} {76}},\ \bibinfo {pages} {769} (\bibinfo {year} {1949})}\BibitemShut
  {NoStop}%
\bibitem [{\citenamefont {Schwinger}(1948{\natexlab{b}})}]{Schwinger:1948iu}%
  \BibitemOpen
  \bibfield  {author} {\bibinfo {author} {\bibfnamefont {J.~S.}\ \bibnamefont
  {Schwinger}},\ }\bibfield  {title} {\bibinfo {title} {{On Quantum
  electrodynamics and the magnetic moment of the electron}},\ }\href
  {https://doi.org/10.1103/PhysRev.73.416} {\bibfield  {journal} {\bibinfo
  {journal} {Phys. Rev.}\ }\textbf {\bibinfo {volume} {73}},\ \bibinfo {pages}
  {416} (\bibinfo {year} {1948}{\natexlab{b}})}\BibitemShut {NoStop}%
\bibitem [{\citenamefont {Petermann}(1957)}]{Petermann:1957hs}%
  \BibitemOpen
  \bibfield  {author} {\bibinfo {author} {\bibfnamefont {A.}~\bibnamefont
  {Petermann}},\ }\bibfield  {title} {\bibinfo {title} {{Fourth order magnetic
  moment of the electron}},\ }\href {https://doi.org/10.5169/seals-112823}
  {\bibfield  {journal} {\bibinfo  {journal} {Helv. Phys. Acta}\ }\textbf
  {\bibinfo {volume} {30}},\ \bibinfo {pages} {407} (\bibinfo {year}
  {1957})}\BibitemShut {NoStop}%
\bibitem [{\citenamefont {Sommerfield}(1957)}]{Sommerfield:1957zz}%
  \BibitemOpen
  \bibfield  {author} {\bibinfo {author} {\bibfnamefont {C.~M.}\ \bibnamefont
  {Sommerfield}},\ }\bibfield  {title} {\bibinfo {title} {{Magnetic Dipole
  Moment of the Electron}},\ }\href {https://doi.org/10.1103/PhysRev.107.328}
  {\bibfield  {journal} {\bibinfo  {journal} {Phys. Rev.}\ }\textbf {\bibinfo
  {volume} {107}},\ \bibinfo {pages} {328} (\bibinfo {year}
  {1957})}\BibitemShut {NoStop}%
\bibitem [{\citenamefont {Sommerfield}(1958)}]{Sommerfield:1958asm}%
  \BibitemOpen
  \bibfield  {author} {\bibinfo {author} {\bibfnamefont {C.~M.}\ \bibnamefont
  {Sommerfield}},\ }\bibfield  {title} {\bibinfo {title} {{The magnetic moment
  of the electron}},\ }\href {https://doi.org/10.1016/0003-4916(58)90003-4}
  {\bibfield  {journal} {\bibinfo  {journal} {Annals Phys.}\ }\textbf {\bibinfo
  {volume} {5}},\ \bibinfo {pages} {26} (\bibinfo {year} {1958})}\BibitemShut
  {NoStop}%
\bibitem [{\citenamefont {Laporta}(2017)}]{Laporta:2017okg}%
  \BibitemOpen
  \bibfield  {author} {\bibinfo {author} {\bibfnamefont {S.}~\bibnamefont
  {Laporta}},\ }\bibfield  {title} {\bibinfo {title} {{High-precision
  calculation of the 4-loop contribution to the electron g-2 in QED}},\ }\href
  {https://doi.org/10.1016/j.physletb.2017.06.056} {\bibfield  {journal}
  {\bibinfo  {journal} {Phys. Lett. B}\ }\textbf {\bibinfo {volume} {772}},\
  \bibinfo {pages} {232} (\bibinfo {year} {2017})},\ \Eprint
  {https://arxiv.org/abs/1704.06996} {arXiv:1704.06996 [hep-ph]} \BibitemShut
  {NoStop}%
\bibitem [{\citenamefont {Fan}\ \emph {et~al.}(2023)\citenamefont {Fan},
  \citenamefont {Myers}, \citenamefont {Sukra},\ and\ \citenamefont
  {Gabrielse}}]{Fan:2023edw}%
  \BibitemOpen
  \bibfield  {author} {\bibinfo {author} {\bibfnamefont {X.}~\bibnamefont
  {Fan}}, \bibinfo {author} {\bibfnamefont {T.~G.}\ \bibnamefont {Myers}},
  \bibinfo {author} {\bibfnamefont {B.~A.~D.}\ \bibnamefont {Sukra}},\ and\
  \bibinfo {author} {\bibfnamefont {G.}~\bibnamefont {Gabrielse}},\ }\bibfield
  {title} {\bibinfo {title} {{Measurement of the Electron Magnetic Moment}},\
  }\href {https://doi.org/10.1103/PhysRevLett.130.071801} {\bibfield  {journal}
  {\bibinfo  {journal} {Phys. Rev. Lett.}\ }\textbf {\bibinfo {volume} {130}},\
  \bibinfo {pages} {071801} (\bibinfo {year} {2023})},\ \Eprint
  {https://arxiv.org/abs/2209.13084} {arXiv:2209.13084 [physics.atom-ph]}
  \BibitemShut {NoStop}%
\bibitem [{\citenamefont {Aliberti}\ \emph {et~al.}(2025)\citenamefont
  {Aliberti} \emph {et~al.}}]{Aliberti:2025beg}%
  \BibitemOpen
  \bibfield  {author} {\bibinfo {author} {\bibfnamefont {R.}~\bibnamefont
  {Aliberti}} \emph {et~al.},\ }\bibfield  {title} {\bibinfo {title} {{The
  anomalous magnetic moment of the muon in the Standard Model: an update}},\
  }\href {https://doi.org/10.1016/j.physrep.2025.08.002} {\bibfield  {journal}
  {\bibinfo  {journal} {Phys. Rept.}\ }\textbf {\bibinfo {volume} {1143}},\
  \bibinfo {pages} {1} (\bibinfo {year} {2025})},\ \Eprint
  {https://arxiv.org/abs/2505.21476} {arXiv:2505.21476 [hep-ph]} \BibitemShut
  {NoStop}%
\bibitem [{\citenamefont {Aoyama}\ \emph {et~al.}(2015)\citenamefont {Aoyama},
  \citenamefont {Hayakawa}, \citenamefont {Kinoshita},\ and\ \citenamefont
  {Nio}}]{Aoyama:2014sxa}%
  \BibitemOpen
  \bibfield  {author} {\bibinfo {author} {\bibfnamefont {T.}~\bibnamefont
  {Aoyama}}, \bibinfo {author} {\bibfnamefont {M.}~\bibnamefont {Hayakawa}},
  \bibinfo {author} {\bibfnamefont {T.}~\bibnamefont {Kinoshita}},\ and\
  \bibinfo {author} {\bibfnamefont {M.}~\bibnamefont {Nio}},\ }\bibfield
  {title} {\bibinfo {title} {{Tenth-Order Electron Anomalous Magnetic Moment
  --- Contribution of Diagrams without Closed Lepton Loops}},\ }\href
  {https://doi.org/10.1103/PhysRevD.91.033006} {\bibfield  {journal} {\bibinfo
  {journal} {Phys. Rev. D}\ }\textbf {\bibinfo {volume} {91}},\ \bibinfo
  {pages} {033006} (\bibinfo {year} {2015})},\ \bibinfo {note} {[Erratum:
  Phys.Rev.D 96, 019901 (2017)]},\ \Eprint {https://arxiv.org/abs/1412.8284}
  {arXiv:1412.8284 [hep-ph]} \BibitemShut {NoStop}%
\bibitem [{\citenamefont {Aoyama}\ \emph
  {et~al.}(2012{\natexlab{a}})\citenamefont {Aoyama}, \citenamefont {Hayakawa},
  \citenamefont {Kinoshita},\ and\ \citenamefont {Nio}}]{Aoyama:2012wj}%
  \BibitemOpen
  \bibfield  {author} {\bibinfo {author} {\bibfnamefont {T.}~\bibnamefont
  {Aoyama}}, \bibinfo {author} {\bibfnamefont {M.}~\bibnamefont {Hayakawa}},
  \bibinfo {author} {\bibfnamefont {T.}~\bibnamefont {Kinoshita}},\ and\
  \bibinfo {author} {\bibfnamefont {M.}~\bibnamefont {Nio}},\ }\bibfield
  {title} {\bibinfo {title} {{Tenth-Order QED Contribution to the Electron g-2
  and an Improved Value of the Fine Structure Constant}},\ }\href
  {https://doi.org/10.1103/PhysRevLett.109.111807} {\bibfield  {journal}
  {\bibinfo  {journal} {Phys. Rev. Lett.}\ }\textbf {\bibinfo {volume} {109}},\
  \bibinfo {pages} {111807} (\bibinfo {year} {2012}{\natexlab{a}})},\ \Eprint
  {https://arxiv.org/abs/1205.5368} {arXiv:1205.5368 [hep-ph]} \BibitemShut
  {NoStop}%
\bibitem [{\citenamefont {Aoyama}\ \emph
  {et~al.}(2012{\natexlab{b}})\citenamefont {Aoyama}, \citenamefont {Hayakawa},
  \citenamefont {Kinoshita},\ and\ \citenamefont {Nio}}]{Aoyama:2012wk}%
  \BibitemOpen
  \bibfield  {author} {\bibinfo {author} {\bibfnamefont {T.}~\bibnamefont
  {Aoyama}}, \bibinfo {author} {\bibfnamefont {M.}~\bibnamefont {Hayakawa}},
  \bibinfo {author} {\bibfnamefont {T.}~\bibnamefont {Kinoshita}},\ and\
  \bibinfo {author} {\bibfnamefont {M.}~\bibnamefont {Nio}},\ }\bibfield
  {title} {\bibinfo {title} {{Complete Tenth-Order QED Contribution to the Muon
  g-2}},\ }\href {https://doi.org/10.1103/PhysRevLett.109.111808} {\bibfield
  {journal} {\bibinfo  {journal} {Phys. Rev. Lett.}\ }\textbf {\bibinfo
  {volume} {109}},\ \bibinfo {pages} {111808} (\bibinfo {year}
  {2012}{\natexlab{b}})},\ \Eprint {https://arxiv.org/abs/1205.5370}
  {arXiv:1205.5370 [hep-ph]} \BibitemShut {NoStop}%
\bibitem [{\citenamefont {Gabrielse}\ \emph {et~al.}(2006)\citenamefont
  {Gabrielse}, \citenamefont {Hanneke}, \citenamefont {Kinoshita},
  \citenamefont {Nio},\ and\ \citenamefont {Odom}}]{Gabrielse:2006gg}%
  \BibitemOpen
  \bibfield  {author} {\bibinfo {author} {\bibfnamefont {G.}~\bibnamefont
  {Gabrielse}}, \bibinfo {author} {\bibfnamefont {D.}~\bibnamefont {Hanneke}},
  \bibinfo {author} {\bibfnamefont {T.}~\bibnamefont {Kinoshita}}, \bibinfo
  {author} {\bibfnamefont {M.}~\bibnamefont {Nio}},\ and\ \bibinfo {author}
  {\bibfnamefont {B.~C.}\ \bibnamefont {Odom}},\ }\bibfield  {title} {\bibinfo
  {title} {{New Determination of the Fine Structure Constant from the Electron
  g Value and QED}},\ }\href {https://doi.org/10.1103/PhysRevLett.97.030802}
  {\bibfield  {journal} {\bibinfo  {journal} {Phys. Rev. Lett.}\ }\textbf
  {\bibinfo {volume} {97}},\ \bibinfo {pages} {030802} (\bibinfo {year}
  {2006})},\ \bibinfo {note} {[Erratum: Phys.Rev.Lett. 99, 039902
  (2007)]}\BibitemShut {NoStop}%
\bibitem [{\citenamefont {Aoyama}\ \emph {et~al.}(2018)\citenamefont {Aoyama},
  \citenamefont {Kinoshita},\ and\ \citenamefont {Nio}}]{Aoyama:2017uqe}%
  \BibitemOpen
  \bibfield  {author} {\bibinfo {author} {\bibfnamefont {T.}~\bibnamefont
  {Aoyama}}, \bibinfo {author} {\bibfnamefont {T.}~\bibnamefont {Kinoshita}},\
  and\ \bibinfo {author} {\bibfnamefont {M.}~\bibnamefont {Nio}},\ }\bibfield
  {title} {\bibinfo {title} {{Revised and Improved Value of the QED Tenth-Order
  Electron Anomalous Magnetic Moment}},\ }\href
  {https://doi.org/10.1103/PhysRevD.97.036001} {\bibfield  {journal} {\bibinfo
  {journal} {Phys. Rev. D}\ }\textbf {\bibinfo {volume} {97}},\ \bibinfo
  {pages} {036001} (\bibinfo {year} {2018})},\ \Eprint
  {https://arxiv.org/abs/1712.06060} {arXiv:1712.06060 [hep-ph]} \BibitemShut
  {NoStop}%
\bibitem [{\citenamefont {Aoyama}\ \emph {et~al.}(2020)\citenamefont {Aoyama}
  \emph {et~al.}}]{Aoyama:2020ynm}%
  \BibitemOpen
  \bibfield  {author} {\bibinfo {author} {\bibfnamefont {T.}~\bibnamefont
  {Aoyama}} \emph {et~al.},\ }\bibfield  {title} {\bibinfo {title} {{The
  anomalous magnetic moment of the muon in the Standard Model}},\ }\href
  {https://doi.org/10.1016/j.physrep.2020.07.006} {\bibfield  {journal}
  {\bibinfo  {journal} {Phys. Rept.}\ }\textbf {\bibinfo {volume} {887}},\
  \bibinfo {pages} {1} (\bibinfo {year} {2020})},\ \Eprint
  {https://arxiv.org/abs/2006.04822} {arXiv:2006.04822 [hep-ph]} \BibitemShut
  {NoStop}%
\bibitem [{\citenamefont {Aoyama}\ \emph {et~al.}(2019)\citenamefont {Aoyama},
  \citenamefont {Kinoshita},\ and\ \citenamefont {Nio}}]{Aoyama:2019ryr}%
  \BibitemOpen
  \bibfield  {author} {\bibinfo {author} {\bibfnamefont {T.}~\bibnamefont
  {Aoyama}}, \bibinfo {author} {\bibfnamefont {T.}~\bibnamefont {Kinoshita}},\
  and\ \bibinfo {author} {\bibfnamefont {M.}~\bibnamefont {Nio}},\ }\bibfield
  {title} {\bibinfo {title} {{Theory of the Anomalous Magnetic Moment of the
  Electron}},\ }\href {https://doi.org/10.3390/atoms7010028} {\bibfield
  {journal} {\bibinfo  {journal} {Atoms}\ }\textbf {\bibinfo {volume} {7}},\
  \bibinfo {pages} {28} (\bibinfo {year} {2019})}\BibitemShut {NoStop}%
\bibitem [{\citenamefont {Volkov}(2016)}]{Volkov:2015pca}%
  \BibitemOpen
  \bibfield  {author} {\bibinfo {author} {\bibfnamefont {S.~A.}\ \bibnamefont
  {Volkov}},\ }\bibfield  {title} {\bibinfo {title} {{Subtractive procedure for
  calculating the anomalous electron magnetic moment in QED and its application
  for numerical calculation at the three-loop level}},\ }\href
  {https://doi.org/10.1134/S1063776116050113} {\bibfield  {journal} {\bibinfo
  {journal} {J. Exp. Theor. Phys.}\ }\textbf {\bibinfo {volume} {122}},\
  \bibinfo {pages} {1008} (\bibinfo {year} {2016})},\ \Eprint
  {https://arxiv.org/abs/1507.06435} {arXiv:1507.06435 [hep-ph]} \BibitemShut
  {NoStop}%
\bibitem [{\citenamefont {Volkov}(2018)}]{Volkov:2018jhy}%
  \BibitemOpen
  \bibfield  {author} {\bibinfo {author} {\bibfnamefont {S.}~\bibnamefont
  {Volkov}},\ }\bibfield  {title} {\bibinfo {title} {{Numerical calculation of
  high-order QED contributions to the electron anomalous magnetic moment}},\
  }\href {https://doi.org/10.1103/PhysRevD.98.076018} {\bibfield  {journal}
  {\bibinfo  {journal} {Phys. Rev. D}\ }\textbf {\bibinfo {volume} {98}},\
  \bibinfo {pages} {076018} (\bibinfo {year} {2018})},\ \Eprint
  {https://arxiv.org/abs/1807.05281} {arXiv:1807.05281 [hep-ph]} \BibitemShut
  {NoStop}%
\bibitem [{\citenamefont {Volkov}(2019)}]{Volkov:2019phy}%
  \BibitemOpen
  \bibfield  {author} {\bibinfo {author} {\bibfnamefont {S.}~\bibnamefont
  {Volkov}},\ }\bibfield  {title} {\bibinfo {title} {{Calculating the five-loop
  QED contribution to the electron anomalous magnetic moment: Graphs without
  lepton loops}},\ }\href {https://doi.org/10.1103/PhysRevD.100.096004}
  {\bibfield  {journal} {\bibinfo  {journal} {Phys. Rev. D}\ }\textbf {\bibinfo
  {volume} {100}},\ \bibinfo {pages} {096004} (\bibinfo {year} {2019})},\
  \Eprint {https://arxiv.org/abs/1909.08015} {arXiv:1909.08015 [hep-ph]}
  \BibitemShut {NoStop}%
\bibitem [{\citenamefont {Lepage}(1978)}]{Lepage:1977sw}%
  \BibitemOpen
  \bibfield  {author} {\bibinfo {author} {\bibfnamefont {G.~P.}\ \bibnamefont
  {Lepage}},\ }\bibfield  {title} {\bibinfo {title} {{A New Algorithm for
  Adaptive Multidimensional Integration}},\ }\href
  {https://doi.org/10.1016/0021-9991(78)90004-9} {\bibfield  {journal}
  {\bibinfo  {journal} {J. Comput. Phys.}\ }\textbf {\bibinfo {volume} {27}},\
  \bibinfo {pages} {192} (\bibinfo {year} {1978})}\BibitemShut {NoStop}%
\bibitem [{\citenamefont {Stenger}(1993)}]{stenger1993numerical}%
  \BibitemOpen
  \bibfield  {author} {\bibinfo {author} {\bibfnamefont {F.}~\bibnamefont
  {Stenger}},\ }\href {https://doi.org/10.1007/978-1-4612-2706-9} {\emph
  {\bibinfo {title} {Numerical Methods Based on Sinc and Analytic
  Functions}}},\ \bibinfo {series} {Springer Series in Computational
  Mathematics}, Vol.~\bibinfo {volume} {20}\ (\bibinfo  {publisher}
  {Springer-Verlag},\ \bibinfo {address} {New York},\ \bibinfo {year}
  {1993})\BibitemShut {NoStop}%
\bibitem [{\citenamefont {Easther}\ \emph {et~al.}(2000)\citenamefont
  {Easther}, \citenamefont {Guralnik},\ and\ \citenamefont
  {Hahn}}]{Easther:1999qb}%
  \BibitemOpen
  \bibfield  {author} {\bibinfo {author} {\bibfnamefont {R.}~\bibnamefont
  {Easther}}, \bibinfo {author} {\bibfnamefont {G.}~\bibnamefont {Guralnik}},\
  and\ \bibinfo {author} {\bibfnamefont {S.}~\bibnamefont {Hahn}},\ }\bibfield
  {title} {\bibinfo {title} {{Fast evaluation of Feynman diagrams}},\ }\href
  {https://doi.org/10.1103/PhysRevD.61.125001} {\bibfield  {journal} {\bibinfo
  {journal} {Phys. Rev. D}\ }\textbf {\bibinfo {volume} {61}},\ \bibinfo
  {pages} {125001} (\bibinfo {year} {2000})},\ \Eprint
  {https://arxiv.org/abs/hep-ph/9903255} {arXiv:hep-ph/9903255} \BibitemShut
  {NoStop}%
\bibitem [{\citenamefont {Easther}\ \emph {et~al.}(2001)\citenamefont
  {Easther}, \citenamefont {Guralnik},\ and\ \citenamefont
  {Hahn}}]{Easther:1999bs}%
  \BibitemOpen
  \bibfield  {author} {\bibinfo {author} {\bibfnamefont {R.}~\bibnamefont
  {Easther}}, \bibinfo {author} {\bibfnamefont {G.}~\bibnamefont {Guralnik}},\
  and\ \bibinfo {author} {\bibfnamefont {S.}~\bibnamefont {Hahn}},\ }\bibfield
  {title} {\bibinfo {title} {{The Sinc function representation and three loop
  master diagrams}},\ }\href {https://doi.org/10.1103/PhysRevD.63.085017}
  {\bibfield  {journal} {\bibinfo  {journal} {Phys. Rev. D}\ }\textbf {\bibinfo
  {volume} {63}},\ \bibinfo {pages} {085017} (\bibinfo {year} {2001})},\
  \Eprint {https://arxiv.org/abs/hep-ph/9912255} {arXiv:hep-ph/9912255}
  \BibitemShut {NoStop}%
\bibitem [{\citenamefont {Petrov}\ \emph {et~al.}(2001)\citenamefont {Petrov},
  \citenamefont {Easther}, \citenamefont {Guralnik}, \citenamefont {Hahn},\
  and\ \citenamefont {Wang}}]{Petrov:2000am}%
  \BibitemOpen
  \bibfield  {author} {\bibinfo {author} {\bibfnamefont {D.}~\bibnamefont
  {Petrov}}, \bibinfo {author} {\bibfnamefont {R.}~\bibnamefont {Easther}},
  \bibinfo {author} {\bibfnamefont {G.}~\bibnamefont {Guralnik}}, \bibinfo
  {author} {\bibfnamefont {S.}~\bibnamefont {Hahn}},\ and\ \bibinfo {author}
  {\bibfnamefont {W.-M.}\ \bibnamefont {Wang}},\ }\bibfield  {title} {\bibinfo
  {title} {{Fermions, gauge theories, and the sinc function representation for
  Feynman diagrams}},\ }\href {https://doi.org/10.1103/PhysRevD.63.105001}
  {\bibfield  {journal} {\bibinfo  {journal} {Phys. Rev. D}\ }\textbf {\bibinfo
  {volume} {63}},\ \bibinfo {pages} {105001} (\bibinfo {year} {2001})},\
  \Eprint {https://arxiv.org/abs/hep-ph/0010143} {arXiv:hep-ph/0010143}
  \BibitemShut {NoStop}%
\bibitem [{\citenamefont {Passarino}(2025)}]{Passarino:2025cxi}%
  \BibitemOpen
  \bibfield  {author} {\bibinfo {author} {\bibfnamefont {G.}~\bibnamefont
  {Passarino}},\ }\bibfield  {title} {\bibinfo {title} {{Numerical computation
  of Fox functions}},\ }\href@noop {} {\  (\bibinfo {year} {2025})},\ \Eprint
  {https://arxiv.org/abs/2506.09597} {arXiv:2506.09597 [hep-ph]} \BibitemShut
  {NoStop}%
\bibitem [{\citenamefont {Slepian}(2021)}]{Slepian:2018vds}%
  \BibitemOpen
  \bibfield  {author} {\bibinfo {author} {\bibfnamefont {Z.}~\bibnamefont
  {Slepian}},\ }\bibfield  {title} {\bibinfo {title} {{On decoupling the
  integrals of cosmological perturbation theory}},\ }\href
  {https://doi.org/10.1093/mnras/staa1789} {\bibfield  {journal} {\bibinfo
  {journal} {Mon. Not. Roy. Astron. Soc.}\ }\textbf {\bibinfo {volume} {507}},\
  \bibinfo {pages} {1337} (\bibinfo {year} {2021})},\ \Eprint
  {https://arxiv.org/abs/1812.02728} {arXiv:1812.02728 [astro-ph.CO]}
  \BibitemShut {NoStop}%
\bibitem [{\citenamefont {Bogoliubov}\ and\ \citenamefont
  {Parasiuk}(1957)}]{Bogoliubov:1957gp}%
  \BibitemOpen
  \bibfield  {author} {\bibinfo {author} {\bibfnamefont {N.~N.}\ \bibnamefont
  {Bogoliubov}}\ and\ \bibinfo {author} {\bibfnamefont {O.~S.}\ \bibnamefont
  {Parasiuk}},\ }\bibfield  {title} {\bibinfo {title} {{On the Multiplication
  of the causal function in the quantum theory of fields}},\ }\href
  {https://doi.org/10.1007/BF02392399} {\bibfield  {journal} {\bibinfo
  {journal} {Acta Math.}\ }\textbf {\bibinfo {volume} {97}},\ \bibinfo {pages}
  {227} (\bibinfo {year} {1957})}\BibitemShut {NoStop}%
\bibitem [{\citenamefont {Hepp}(1966)}]{Hepp:1966eg}%
  \BibitemOpen
  \bibfield  {author} {\bibinfo {author} {\bibfnamefont {K.}~\bibnamefont
  {Hepp}},\ }\bibfield  {title} {\bibinfo {title} {{Proof of the
  Bogolyubov-Parasiuk theorem on renormalization}},\ }\href
  {https://doi.org/10.1007/BF01773358} {\bibfield  {journal} {\bibinfo
  {journal} {Commun. Math. Phys.}\ }\textbf {\bibinfo {volume} {2}},\ \bibinfo
  {pages} {301} (\bibinfo {year} {1966})}\BibitemShut {NoStop}%
\bibitem [{\citenamefont {Zimmermann}(1969)}]{Zimmermann1969}%
  \BibitemOpen
  \bibfield  {author} {\bibinfo {author} {\bibfnamefont {W.}~\bibnamefont
  {Zimmermann}},\ }\bibfield  {title} {\bibinfo {title} {Convergence of
  {Bogolyubov's} method of renormalization in momentum space},\ }\href
  {https://doi.org/10.1007/BF01645676} {\bibfield  {journal} {\bibinfo
  {journal} {Commun. Math. Phys.}\ }\textbf {\bibinfo {volume} {15}},\ \bibinfo
  {pages} {208} (\bibinfo {year} {1969})}\BibitemShut {NoStop}%
\bibitem [{\citenamefont {Laporta}\ and\ \citenamefont
  {Remiddi}(1996)}]{Laporta:1996mq}%
  \BibitemOpen
  \bibfield  {author} {\bibinfo {author} {\bibfnamefont {S.}~\bibnamefont
  {Laporta}}\ and\ \bibinfo {author} {\bibfnamefont {E.}~\bibnamefont
  {Remiddi}},\ }\bibfield  {title} {\bibinfo {title} {{The Analytical value of
  the electron ($g-2$) at order $\alpha^3$ in QED}},\ }\href
  {https://doi.org/10.1016/0370-2693(96)00439-X} {\bibfield  {journal}
  {\bibinfo  {journal} {Phys. Lett. B}\ }\textbf {\bibinfo {volume} {379}},\
  \bibinfo {pages} {283} (\bibinfo {year} {1996})},\ \Eprint
  {https://arxiv.org/abs/hep-ph/9602417} {arXiv:hep-ph/9602417} \BibitemShut
  {NoStop}%
\bibitem [{\citenamefont {Trefethen}\ and\ \citenamefont
  {Weideman}(2014)}]{trefethen2014exponentially}%
  \BibitemOpen
  \bibfield  {author} {\bibinfo {author} {\bibfnamefont {L.~N.}\ \bibnamefont
  {Trefethen}}\ and\ \bibinfo {author} {\bibfnamefont {J.~A.~C.}\ \bibnamefont
  {Weideman}},\ }\bibfield  {title} {\bibinfo {title} {The exponentially
  convergent trapezoidal rule},\ }\href {https://doi.org/10.1137/130932132}
  {\bibfield  {journal} {\bibinfo  {journal} {SIAM Review}\ }\textbf {\bibinfo
  {volume} {56}},\ \bibinfo {pages} {385} (\bibinfo {year} {2014})}\BibitemShut
  {NoStop}%
\bibitem [{\citenamefont {Braess}\ and\ \citenamefont
  {Hackbusch}(2005)}]{Braess:2005ae}%
  \BibitemOpen
  \bibfield  {author} {\bibinfo {author} {\bibfnamefont {D.}~\bibnamefont
  {Braess}}\ and\ \bibinfo {author} {\bibfnamefont {W.}~\bibnamefont
  {Hackbusch}},\ }\bibfield  {title} {\bibinfo {title} {{Approximation of $1/x$
  by exponential sums in $[1,\infty)$}},\ }\href
  {https://doi.org/10.1093/imanum/dri015} {\bibfield  {journal} {\bibinfo
  {journal} {IMA J. Numer. Anal.}\ }\textbf {\bibinfo {volume} {25}},\ \bibinfo
  {pages} {685} (\bibinfo {year} {2005})}\BibitemShut {NoStop}%
\bibitem [{\citenamefont {Beylkin}\ and\ \citenamefont
  {Monz{\'o}n}(2005)}]{Beylkin:2005oa}%
  \BibitemOpen
  \bibfield  {author} {\bibinfo {author} {\bibfnamefont {G.}~\bibnamefont
  {Beylkin}}\ and\ \bibinfo {author} {\bibfnamefont {L.}~\bibnamefont
  {Monz{\'o}n}},\ }\bibfield  {title} {\bibinfo {title} {{On approximation of
  functions by exponential sums}},\ }\href
  {https://doi.org/10.1016/j.acha.2005.01.003} {\bibfield  {journal} {\bibinfo
  {journal} {Appl. Comput. Harmon. Anal.}\ }\textbf {\bibinfo {volume} {19}},\
  \bibinfo {pages} {17} (\bibinfo {year} {2005})}\BibitemShut {NoStop}%
\bibitem [{\citenamefont {Exl}\ \emph {et~al.}(2016)\citenamefont {Exl},
  \citenamefont {Mauser},\ and\ \citenamefont {Zhang}}]{Exl:2016ae}%
  \BibitemOpen
  \bibfield  {author} {\bibinfo {author} {\bibfnamefont {L.}~\bibnamefont
  {Exl}}, \bibinfo {author} {\bibfnamefont {N.~J.}\ \bibnamefont {Mauser}},\
  and\ \bibinfo {author} {\bibfnamefont {Y.}~\bibnamefont {Zhang}},\ }\bibfield
   {title} {\bibinfo {title} {{Accurate and efficient computation of nonlocal
  potentials based on Gaussian-sum approximation}},\ }\href
  {https://doi.org/10.1016/j.jcp.2016.09.045} {\bibfield  {journal} {\bibinfo
  {journal} {J. Comput. Phys.}\ }\textbf {\bibinfo {volume} {327}},\ \bibinfo
  {pages} {629} (\bibinfo {year} {2016})},\ \Eprint
  {https://arxiv.org/abs/1501.04438} {arXiv:1501.04438 [physics.comp-ph]}
  \BibitemShut {NoStop}%
\bibitem [{\citenamefont {Easther}(2026)}]{sincFeyn}%
  \BibitemOpen
  \bibfield  {author} {\bibinfo {author} {\bibfnamefont {R.}~\bibnamefont
  {Easther}},\ }\href@noop {} {\bibinfo {title} {{sincFeyn}, version 0.9.0}},\
  \bibinfo {howpublished} {\url{https://github.com/reasther/sincFeyn}}
  (\bibinfo {year} {2026}),\ \bibinfo {note} {{Zenodo},
  \href{https://doi.org/10.5281/zenodo.23975727}{doi:10.5281/zenodo.23975727}}\BibitemShut
  {NoStop}%
\bibitem [{\citenamefont {Peskin}\ and\ \citenamefont
  {Schroeder}(1995)}]{Peskin:1995ev}%
  \BibitemOpen
  \bibfield  {author} {\bibinfo {author} {\bibfnamefont {M.~E.}\ \bibnamefont
  {Peskin}}\ and\ \bibinfo {author} {\bibfnamefont {D.~V.}\ \bibnamefont
  {Schroeder}},\ }\href {https://doi.org/10.1201/9780429503559} {\emph
  {\bibinfo {title} {{An Introduction to quantum field theory}}}}\ (\bibinfo
  {publisher} {Addison-Wesley},\ \bibinfo {address} {Reading, USA},\ \bibinfo
  {year} {1995})\BibitemShut {NoStop}%
\bibitem [{\citenamefont {Volkov}(2017)}]{Volkov:2017xaq}%
  \BibitemOpen
  \bibfield  {author} {\bibinfo {author} {\bibfnamefont {S.}~\bibnamefont
  {Volkov}},\ }\bibfield  {title} {\bibinfo {title} {{New method of computing
  the contributions of graphs without lepton loops to the electron anomalous
  magnetic moment in QED}},\ }\href
  {https://doi.org/10.1103/PhysRevD.96.096018} {\bibfield  {journal} {\bibinfo
  {journal} {Phys. Rev. D}\ }\textbf {\bibinfo {volume} {96}},\ \bibinfo
  {pages} {096018} (\bibinfo {year} {2017})},\ \Eprint
  {https://arxiv.org/abs/1705.05800} {arXiv:1705.05800 [hep-ph]} \BibitemShut
  {NoStop}%
\bibitem [{\citenamefont {Mallat}(2008)}]{mallat2008wavelet}%
  \BibitemOpen
  \bibfield  {author} {\bibinfo {author} {\bibfnamefont {S.}~\bibnamefont
  {Mallat}},\ }\href@noop {} {\emph {\bibinfo {title} {A Wavelet Tour of Signal
  Processing: The Sparse Way}}},\ \bibinfo {edition} {3rd}\ ed.\ (\bibinfo
  {publisher} {Academic Press},\ \bibinfo {address} {Burlington, MA},\ \bibinfo
  {year} {2008})\BibitemShut {NoStop}%
\bibitem [{\citenamefont {Cheng}\ and\ \citenamefont
  {Wu}(1987)}]{Cheng1987Expanding}%
  \BibitemOpen
  \bibfield  {author} {\bibinfo {author} {\bibfnamefont {H.}~\bibnamefont
  {Cheng}}\ and\ \bibinfo {author} {\bibfnamefont {T.~T.}\ \bibnamefont {Wu}},\
  }\href@noop {} {\emph {\bibinfo {title} {Expanding Protons: Scattering at
  High Energies}}}\ (\bibinfo  {publisher} {The MIT Press},\ \bibinfo {address}
  {Cambridge, MA},\ \bibinfo {year} {1987})\ p.\ \bibinfo {pages}
  {285}\BibitemShut {NoStop}%
\bibitem [{\citenamefont {Davies}\ \emph {et~al.}(2026)\citenamefont {Davies},
  \citenamefont {Kaneko}, \citenamefont {Marinissen}, \citenamefont {Ueda},\
  and\ \citenamefont {Vermaseren}}]{Davies:2026cci}%
  \BibitemOpen
  \bibfield  {author} {\bibinfo {author} {\bibfnamefont {J.}~\bibnamefont
  {Davies}}, \bibinfo {author} {\bibfnamefont {T.}~\bibnamefont {Kaneko}},
  \bibinfo {author} {\bibfnamefont {C.}~\bibnamefont {Marinissen}}, \bibinfo
  {author} {\bibfnamefont {T.}~\bibnamefont {Ueda}},\ and\ \bibinfo {author}
  {\bibfnamefont {J.~A.~M.}\ \bibnamefont {Vermaseren}},\ }\bibfield  {title}
  {\bibinfo {title} {Form version 5.0},\ }\href@noop {} {\  (\bibinfo {year}
  {2026})},\ \Eprint {https://arxiv.org/abs/2601.19982} {arXiv:2601.19982
  [hep-ph]} \BibitemShut {NoStop}%
\bibitem [{\citenamefont {Ferguson}\ \emph {et~al.}(1999)\citenamefont
  {Ferguson}, \citenamefont {Bailey},\ and\ \citenamefont
  {Arno}}]{Ferguson:1999pslq}%
  \BibitemOpen
  \bibfield  {author} {\bibinfo {author} {\bibfnamefont {H.~R.~P.}\
  \bibnamefont {Ferguson}}, \bibinfo {author} {\bibfnamefont {D.~H.}\
  \bibnamefont {Bailey}},\ and\ \bibinfo {author} {\bibfnamefont
  {S.}~\bibnamefont {Arno}},\ }\bibfield  {title} {\bibinfo {title} {Analysis
  of {PSLQ}, an integer relation finding algorithm},\ }\href@noop {} {\bibfield
   {journal} {\bibinfo  {journal} {Math. Comp.}\ }\textbf {\bibinfo {volume}
  {68}},\ \bibinfo {pages} {351} (\bibinfo {year} {1999})}\BibitemShut
  {NoStop}%
\bibitem [{\citenamefont {{The mpmath development team}}(2023)}]{mpmath}%
  \BibitemOpen
  \bibfield  {author} {\bibinfo {author} {\bibnamefont {{The mpmath development
  team}}},\ }\href@noop {} {\emph {\bibinfo {title} {mpmath: a {P}ython library
  for arbitrary-precision floating-point arithmetic (version 1.3.0)}}}
  (\bibinfo {year} {2023}),\ \bibinfo {note}
  {\url{https://mpmath.org/}}\BibitemShut {NoStop}%
\bibitem [{\citenamefont {{NVIDIA Corporation}}(2014)}]{nvidia2014teslak80}%
  \BibitemOpen
  \bibfield  {author} {\bibinfo {author} {\bibnamefont {{NVIDIA
  Corporation}}},\ }\href@noop {} {\emph {\bibinfo {title} {{NVIDIA Tesla K80
  GPU Accelerator Board Spec}}}},\ \bibinfo {type} {Tech. Rep.}\ (\bibinfo
  {institution} {NVIDIA Corporation},\ \bibinfo {year} {2014})\ \bibinfo {note}
  {dual-GPU GK210 accelerator, 24 GB GDDR5, 2.91 TFLOPS FP64 peak performance
  at boost clock}\BibitemShut {NoStop}%
\bibitem [{\citenamefont {{NVIDIA Corporation}}(2017)}]{nvidia2017teslav100}%
  \BibitemOpen
  \bibfield  {author} {\bibinfo {author} {\bibnamefont {{NVIDIA
  Corporation}}},\ }\href@noop {} {\emph {\bibinfo {title} {{NVIDIA Tesla V100
  GPU Architecture}}}},\ \bibinfo {type} {Tech. Rep.}\ \bibinfo {number}
  {WP-08608-001}\ (\bibinfo  {institution} {NVIDIA Corporation},\ \bibinfo
  {year} {2017})\ \bibinfo {note} {whitepaper. FP64: 7.8 TFLOPS (SXM2), 7.0
  TFLOPS (PCIe)}\BibitemShut {NoStop}%
\bibitem [{\citenamefont {{NVIDIA Corporation}}(2020)}]{nvidia2020a100}%
  \BibitemOpen
  \bibfield  {author} {\bibinfo {author} {\bibnamefont {{NVIDIA
  Corporation}}},\ }\href
  {https://images.nvidia.com/aem-dam/en-zz/Solutions/data-center/nvidia-ampere-architecture-whitepaper.pdf}
  {\emph {\bibinfo {title} {NVIDIA A100 Tensor Core GPU Architecture}}},\
  \bibinfo {type} {Whitepaper}\ (\bibinfo  {institution} {NVIDIA Corporation},\
  \bibinfo {year} {2020})\BibitemShut {NoStop}%
\bibitem [{\citenamefont {Volkov}(2024)}]{Volkov:2024yzc}%
  \BibitemOpen
  \bibfield  {author} {\bibinfo {author} {\bibfnamefont {S.}~\bibnamefont
  {Volkov}},\ }\bibfield  {title} {\bibinfo {title} {{Calculation of the total
  10th order QED contribution to the electron magnetic moment}},\ }\href
  {https://doi.org/10.1103/PhysRevD.110.036001} {\bibfield  {journal} {\bibinfo
   {journal} {Phys. Rev. D}\ }\textbf {\bibinfo {volume} {110}},\ \bibinfo
  {pages} {036001} (\bibinfo {year} {2024})},\ \Eprint
  {https://arxiv.org/abs/2404.00649} {arXiv:2404.00649 [hep-ph]} \BibitemShut
  {NoStop}%
\bibitem [{\citenamefont {Aoyama}\ \emph {et~al.}(2025)\citenamefont {Aoyama},
  \citenamefont {Hayakawa}, \citenamefont {Hirayama},\ and\ \citenamefont
  {Nio}}]{Aoyama:2024aly}%
  \BibitemOpen
  \bibfield  {author} {\bibinfo {author} {\bibfnamefont {T.}~\bibnamefont
  {Aoyama}}, \bibinfo {author} {\bibfnamefont {M.}~\bibnamefont {Hayakawa}},
  \bibinfo {author} {\bibfnamefont {A.}~\bibnamefont {Hirayama}},\ and\
  \bibinfo {author} {\bibfnamefont {M.}~\bibnamefont {Nio}},\ }\bibfield
  {title} {\bibinfo {title} {{Verification of the tenth-order QED contribution
  to the anomalous magnetic moment of the electron from diagrams without
  fermion loops}},\ }\href {https://doi.org/10.1103/PhysRevD.111.L031902}
  {\bibfield  {journal} {\bibinfo  {journal} {Phys. Rev. D}\ }\textbf {\bibinfo
  {volume} {111}},\ \bibinfo {pages} {L031902} (\bibinfo {year} {2025})},\
  \Eprint {https://arxiv.org/abs/2412.06473} {arXiv:2412.06473 [hep-ph]}
  \BibitemShut {NoStop}%
\bibitem [{\citenamefont {Borowka}\ \emph {et~al.}(2018)\citenamefont
  {Borowka}, \citenamefont {Heinrich}, \citenamefont {Jahn}, \citenamefont
  {Jones}, \citenamefont {Kerner},\ and\ \citenamefont
  {Schlenk}}]{Borowka:2017smv}%
  \BibitemOpen
  \bibfield  {author} {\bibinfo {author} {\bibfnamefont {S.}~\bibnamefont
  {Borowka}}, \bibinfo {author} {\bibfnamefont {G.}~\bibnamefont {Heinrich}},
  \bibinfo {author} {\bibfnamefont {S.~P.}\ \bibnamefont {Jahn}}, \bibinfo
  {author} {\bibfnamefont {S.~P.}\ \bibnamefont {Jones}}, \bibinfo {author}
  {\bibfnamefont {M.}~\bibnamefont {Kerner}},\ and\ \bibinfo {author}
  {\bibfnamefont {J.}~\bibnamefont {Schlenk}},\ }\bibfield  {title} {\bibinfo
  {title} {{pySecDec: a toolbox for the numerical evaluation of multi-scale
  integrals}},\ }\href {https://doi.org/10.1016/j.cpc.2017.09.015} {\bibfield
  {journal} {\bibinfo  {journal} {Comput. Phys. Commun.}\ }\textbf {\bibinfo
  {volume} {222}},\ \bibinfo {pages} {313} (\bibinfo {year} {2018})},\ \Eprint
  {https://arxiv.org/abs/1703.09692} {arXiv:1703.09692 [hep-ph]} \BibitemShut
  {NoStop}%
\end{thebibliography}

\end{document}